\documentclass[journal=langmuir,manuscript=article]{achemso}
\usepackage{graphicx} 
\usepackage{amsmath, amssymb}
\usepackage{times}
\usepackage{xcolor}
\usepackage{blindtext}
\usepackage{hyperref}
\usepackage{comment}
\usepackage[normalem]{ulem}

\title{A Free Energy Model for the Plateau Shear Modulus in Thermosensitive Microgel Suspensions}

\author{Maxime Bergman}
\affiliation{Department of Physics, University of Fribourg, 1700 Fribourg, Switzerland}
\author{Yixuan Xu}
\affiliation{Department of Materials Science and Engineering, University of California - Los Angeles, CA 90095, USA}
\author{Jos\'e Mu\~n\'eton D\'iaz}
\affiliation{Department of Physics, University of Fribourg, 1700 Fribourg, Switzerland}
\author{Chi Zhang}
\affiliation{Department of Physics, University of Fribourg, 1700 Fribourg, Switzerland}
\author{Thomas G. Mason}
\affiliation{Department of Physics and Astronomy, University of California - Los Angeles, CA 90095 , USA}
\alsoaffiliation{Department of Chemistry and Biochemistry, University of California - Los Angeles, CA 90095, USA}
\author{Frank Scheffold}
\affiliation{Department of Physics, University of Fribourg, 1700 Fribourg, Switzerland}
\email{frank.scheffold@unifr.ch}

\begin{document}
\setkeys{acs}{
    keywords = {keyword1, keyword2, keyword3}
}

\begin{abstract}
Polymer microgels exhibit intriguing macroscopic flow properties arising from their unique microscopic structure. Microgel colloids usually comprise a crosslinked polymer network with a radially decaying density profile, resulting in a dense core surrounded by a fuzzy corona. Notably, microgels synthesized from poly(N-isopropyl acrylamide) (PNIPAM) are thermoresponsive, capable of adjusting their size and density profile based on temperature. Above the lower critical solution temperature ($T_\text{LCST} \sim 33$ $^\circ$C), the microgel's polymer network collapses, expulsing water through a reversible process. Conversely, below $33$ $^\circ$C, the microgel's network swells, becoming highly compressible and allowing overpacking to effective volume fractions exceeding one. Under conditions of dense packing, microgels undergo deformation in distinct stages: corona compression and faceting, interpenetration, and finally, isotropic compression. Each stage exhibits a characteristic signature in the dense microgel suspensions' yield stress and elastic modulus. Here, we introduce a model for the linear elastic shear modulus by minimizing a quasi-equilibrium free energy, encompassing all relevant energetic contributions. We validate our model by comparing its predictions to experimental results from oscillatory shear rheology tests on microgel suspensions at different densities and temperatures. Our findings demonstrate that combining macroscopic rheological measurements with the model allows for temperature-dependent characterization of polymer interaction parameters.


\end{abstract}

\maketitle

\maketitle

\section*{INTRODUCTION}In the last few decades, the polymer-colloid duality of microgels has captured the interest of scientists from various fields~\cite{sanson2010synthesis,lyon2012polymer,brijitta2019responsive}. Microgels are cross-linked polymer networks that are typically suspended in water and exhibit properties that lie between those of macroscopic polymer gels and colloids. When the monomer N-isopropylacrylamide (NIPAM) is used, microgels become thermoresponsive, with the network structure changing dramatically when the temperature exceeds the lower critical solution temperature (LCST) of approximately 33 $^\circ$C~\cite{wu2003phase,stieger2004small,mason2005density,reufer2009temperature}. Below the LCST, microgels are soft and deformable, while above the LCST, the network expels water, leading to a denser polymer structure and repulsive or attractive hard sphere-like properties. Due to their tunability at biologically relevant temperatures and ease of synthesis, microgels have become popular in many areas of research. They act as drug carriers~\cite{oh2008development}, viscosity modifiers~\cite{andablo2019microgels}, tunable colloidal deplentants~\cite{sacanna2010lock}, scaffolds in tissue engineering~\cite{gan2010thermogelable,bhattacharjee2016liquid}, and are useful for studying glass transition and jamming phenomena~\cite{zhang2009thermal,ikeda2013disentangling}. However, quantifying their mechanical properties remains challenging due to the combination of polymer physics and colloidal phenomena occurring.

At low concentrations, the colloidal nature of microgels dominates, and the suspension viscosity follows hard-sphere approximations such as the Einstein-Batchelor equation for the suspension's viscosity~\cite{batchelor1977effect,senff1999temperature}. However, microgels are soft and flexible and can be overpacked "beyond" space-filling at higher concentrations, considering the volume fraction occupied by the unperturbed microgel size~\cite{pellet2016glass,conley2017,bouhid2017deswelling,nikolov2020behavior}. As the concentration increases, the microgels come in contact and respond to the increased osmotic pressure exerted by their neighbors. Recent superresolution microscopy studies revealed the different steps of the interaction~\cite{conley2019}. First, the microgels' corona compress, then they weakly interpenetrate, followed by deformation (faceting) of the core, and finally, isotropic compression. This results in a transition from the elastic behavior dominated by colloidal interactions towards properties of homogeneous polymer gels~\cite{menut2012does}.

Previous rheological studies have identified the different interpenetration, deformation, and isotropic compression stages~\cite{cloitre2003structure,scheffold2010,romeo2013elasticity,conley2019,scotti2020}. The modulus $G^\prime_p$ shows a sharp increase during the onset of elasticity, associated with corona compression and interpenetration, followed by a linear increase for effective filling fractions above one where density fluctuations vanish. Several models have been proposed to describe the evolution of $G^\prime_p$ of microgel suspensions in a specific regime~\cite{senff1999temperature,scheffold2010}. Still, few have attempted to model the microgel's elasticity over the entire range of concentrations~\cite{romeo2013elasticity,ghosh2019linear}. Here, we present a framework that explicitly considers microgel jamming, deformation, and compression. Furthermore, our model spans the entire concentration range and globally has only one adjustable parameter per temperature.
\section*{EXPERIMENTAL SECTION}
\subsection*{Microgel synthesis}
N-isopropylacrylamide (NIPAM, 97\%, 7.20 g) and N,N'-Methylenebis(acrylamide) (BIS, 99\%, 0.051524 g) were dissolved in 495 g of H$_2$O in a round-bottom flask. The solution was degassed with nitrogen at room temperature for 30 minutes. The temperature was then increased to 70$^\circ$C, and the reaction mixture was allowed to reach equilibrium for 30 minutes. To initiate the reaction, potassium persulfate (KPS, 99\%, 0.23273 g in 6.5 g H$_2$O) was injected. The reaction proceeded for 4 hours under a nitrogen atmosphere at 70$^\circ$C. Afterward, the resulting solution was filtered using glass wool and subjected to dialysis for 2 weeks to remove any unreacted components. 
The resulting microgels have a cross-linking degree of $\sim$~5 mol \%. The mass concentration of the resulting solution was determined as 1.34 wt $\%$, using drying and weighing. The mass density in wt $\%$ is defined as the ratio of the weight of the dried microgel component to the weight of the total sample.
\subsection*{Dynamic Light Scattering} We diluted a small amount of microgel stock solution (2 $\mu$l) with 5 mM KCl in H$_2$O (200 $\mu$l) and filled it into an NMR tube. The NMR tube was then placed inside a 3D-DLS light scattering spectrometer (LS Instruments, Switzerland), equipped with a laser that emitted light at a wavelength of 660~nm. The spectrometer collected scattered light within the range of angles from 40$^\circ$C to 60$^\circ$C, with a step size of 3$^\circ$C. Measurements were conducted at various temperatures: 20$^\circ$C, 24$^\circ$C, 28$^\circ$C, 30$^\circ$C, 35$^\circ$C, and 40$^\circ$C. For each angle, three measurements were taken, each lasting 60 seconds. Before each measurement, the sample was allowed to reach thermal equilibrium at the target temperature for 30 minutes. The autocorrelation functions obtained from the measurements were subjected to standard second cumulant analysis to determine the hydrodynamic radius.
\subsection*{Viscometry}
To conduct viscosity measurements, a pure solvent sample containing 5 mM KCl was prepared, along with several samples containing different concentrations of microgels (0.06 wt $\%$, 0.1 wt $\%$, 0.3 wt $\%$, 0.6 wt $\%$, and 1.0 wt $\%$) in 5 mM KCl. The measurements were carried out at temperatures of 20$^\circ$C, 22$^\circ$C, 24$^\circ$C, 26$^\circ$C, and 28$^\circ$C using a rolling ball viscometer (Anton Paar, Automated Microviscometer). Before each measurement, the samples were allowed to equilibrate at the respective temperature for 30 minutes.
Ten repetitions of measurements a capillary with a diameter of 1.6 mm under an inclination angle of 40°, and using a ball with a diameter of 1.5 mm. To accurately determine the samples' viscosity, we also measured their mass densities at the same temperatures using a densimeter (Anton Paar, DSA 5000 M). After a 30-minute equilibration period, the density was measured twice. The relative viscosity, denoted as $\eta_\textrm{rel}$, was calculated by dividing the viscosity of each sample by the viscosity of the solvent.
\subsection*{Rheology}
By centrifugation at 15,000 g for 30 minutes, we increased the mass density of the stock solution to 7 wt $\%$. This elevated concentration yielded a dense liquid state at room temperature. Multiple rounds of centrifugation and controlled evaporation were performed at room temperature to obtain samples with higher concentrations. To adjust the salinity of the samples to 5 mM KCl, a specific amount of 1 M KCl electrolyte was added. Rheology measurements were conducted using a dilution series, where after each measurement, the remaining sample was diluted with 5 mM KCl and measured again.
Rheology data was acquired using an Anton Paar rheometer (MCR300) in a cone-plate configuration. The cone had a radius of 25 mm, and the distance between the cone and the plate was 0.053 mm. A solvent trap was utilized during the measurements. Before the rheological measurement, the sample was heated to 30$^\circ$C, and then 0.2 ml was transferred onto the pre-heated plate of the rheometer (maintained at 28$^\circ$C) using a syringe. The sample was pre-sheared for 5 minutes at a shear rate of $\frac{d\gamma}{dt} = 100$ s$^{-1}$. Subsequently, the temperature within the rheometer was gradually decreased from 30$^\circ$C to 20$^\circ$C at a rate of 0.5$^\circ$C every 3 minutes. The values of $G'$ and $G''$ were recorded at an amplitude of $\gamma = 0.1\%$ and an angular frequency of $\omega = 10$ rad/s.  We found that the experimental error bars in the logarithmic plots (Figure~\ref{fig:gpfit}) are smaller than the symbol sizes and they are therefore not shown in the plots. Checks were carried out to ensure these measurements were conducted within the linear regime. We also confirmed that the chosen frequency is within the plateau regime of the frequency-dependent storage modulus, which, as shown earlier (e.g., in Fig. 3a of our reference~\cite{conley2019}), extends up to approximately 50 rad/s. Additionally, it was ensured that the sample reached equilibrium within 3 minutes, and a subsequent temperature cycle from 20$^\circ$C back to 28$^\circ$C was performed to confirm reproducibility.
\section*{THEORY AND MODEL}
\begin{figure*}[ht]
\centering
\includegraphics[trim= 0 240 0 300,width=1\linewidth]{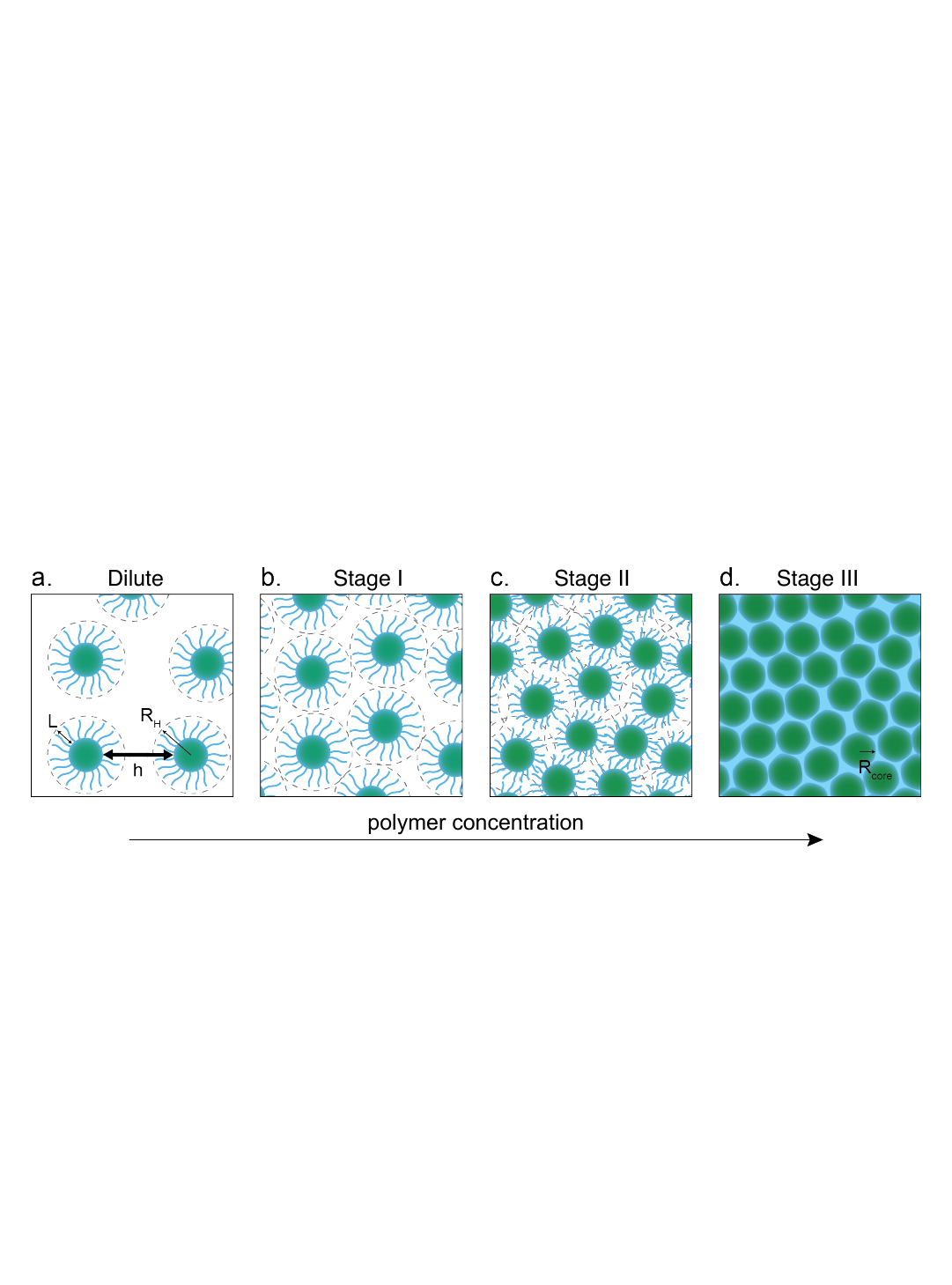}
\caption{Sketch of microgel interactions and deformations with increasing concentration. {\bf{a}} At low effective volume fractions the sample is in liquid state. Increasing volume fraction leads to a critical moment where coronae touch, shown in {\bf{b}}. Dashed lines indicate the overall size of the microgel. {\bf{c}}  Further densification causes the corona to compress. {\bf{d}} Once cores touch, they will deform. Length scales indicated are corona (brush) thickness $L$,  The hydrodynamic radius $ R_\textrm{H}\simeq R$,  where $R$  set the effective volume fraction $\phi$. $h$ denotes the separation distance between cores, and $R_\textrm{core}$ the microgel core radius. }
\label{fig:sketch}
\end{figure*}

Our research aims to establish a link between the various stages of microgel interactions at different temperatures and the distinct elasticity regimes displayed by microgel suspensions and pastes. We consider submicron-sized poly-NIPAM microgels dispersed in water with an ionic strength of 5 mM KCl to screen residual charges carried by the ionic initiators, which is the most commonly studied microgel system. To quantitatively account for the different stages of packing, we propose a free energy minimization scheme. Specifically, we consider the increased number density and the vanishing free volume in stage I, the compression of the microgel corona in stage II, the faceting and compression of the core in stage III (as depicted in Fig.~\ref{fig:sketch}). We exclude the weak interdigitation observed in superresolution microscopy, which affects viscoelastic losses but is not expected to affect the elastic storage modulus substantially~\cite{conley2019}. Here, we argue that interpenetration has a negligible influence on the storage modulus, as the polymer density becomes nearly homogeneous once the corona overlap, as demonstrated by small-angle light scattering in~\cite{conley2017}. Our modeling framework is derived from approach presented by Mason and Scheffold to model the elasticity of dense emulsions that display interfacial deformation~\cite{mason2014crossover}. The model was later expanded to include the double layer repulsion in ionic emulsions by Kim et al.~\cite{kim2016}. Although we choose a similar approach, the interaction potential of microgels differs significantly from that of emulsion droplets~\cite{scheffold2010, conley2017, conley2019}. If Mason and co-workers' model can be extended to encompass microgels, it would introduce a relatively straightforward model capable of explaining the shear modulus in dense microgel suspensions. This would be a notable development, offering a simple tool applicable to various practical uses. 
In our investigation, we aim to assess predictions for the elastic plateau modulus $G^\prime_p$ by comparing them to experimental results obtained from rheometry.
We begin by examining a dilute suspension of microgel particles, characterized by an effective volume fraction $\phi= \frac{N}{V} \frac{4}{3} \pi R^3$. $\frac{N}{V}$ denotes the number density, while $R$ is the effective radius of the microgel particle, which is often assumed to be approximately equal to the hydrodynamic radius $R\simeq R_H$. As shown in Fig.~\ref{fig:sketch} we also consider the core radius $R_\textrm{core}$ and the corresponding core volume fraction $\zeta= \frac{N}{V} \frac{4}{3} \pi R_\textrm{core}^3$. Mason and Scheffold's model has defined the parameter $\zeta_d=\Delta V/V$ as the additional volume fraction that becomes available due to the deformation of the core caused by thermal fluctuations and shear strain~\cite{mason2014crossover}
. Considering a perturbative shear strain amplitude $\gamma$, the sample plateau shear modulus $G^\prime_p$ can be calculated. 
We derive the plateau shear modulus by taking the derivative of the total free energy $F_\textrm{tot}$ under the minimization condition $\zeta_d=\zeta^{*}_d$~\cite{mason2014crossover}. We retain all parameters before evaluating the result at $\gamma\to0$:
\begin{equation}\label{eq:7}
G^\prime_p=|\frac{\delta^2 F_\textrm{tot}}{\delta \gamma^2}|_{\zeta_d=\zeta^{*}_d,\gamma=0}
\end{equation}
\newline  We consider amorphous suspensions; consequently, we expect to find a glass transition at an effective volume fraction of approximately $\phi\simeq0.58$~\cite{bengtzelius1984dynamics,pusey1987observation}.  As the system begins to solidify at the glass transition and beyond, entropic contributions to the free energy dominate due to (transient) particle caging by neighboring particles, leading to arrested motion and increased elasticity up to the critical volume fraction for random close packing (or jamming) of hard spheres where particles touch at approximately $0.646$ volume fraction~\cite{torquato2000random}. 
\begin{table}[ht]
\caption{Parameters for Modeling the Elastic Shear Modulus}
\begin{tabular}{l|l|l|}
 Label &  Meaning &  Determined using \\
 \hline
 $c$ & mass concentration (wt $\%$)   & drying, weighing \\
    $k=\phi/c$  & swelling ratio  & viscosity $\eta\left(\phi\ll 1\right)$  \\
  $\phi= \frac{N}{V}\frac{4}{3} \pi R^3$  &  effective volume fraction & 
 $\phi=k c$  \\
 $R=\left(\frac{V}{N}\phi\right)^\frac{1}{3}$ &  effective radius (nm) &   number density $\frac{N}{V}$  \\
 $k_\text{core}=\zeta/c$  & core swelling ratio & $\frac{G^\prime_p}{\zeta (\zeta-0.646)} = \frac{12}{10} \alpha \left(E^\ast \xi \right)  $ \\ 
   $\zeta= \frac{N}{V}\frac{4}{3} \pi R_\text{core}^3$  &  core volume fraction & 
 $\zeta = k_\text{core} c$  \\
  $R_\text{core}=\left(\frac{V}{N}\zeta\right)^\frac{1}{3}$ &  core radius (nm) &   number density $\frac{N}{V}$  \\
  $L$ & corona thickness (nm) &   $L=R-R_\text{core}$   \\
    $E^\ast $  & core contact modulus (N/m$^2$)& $\frac{G^\prime_p}{\zeta (\zeta-0.646)} = \frac{12}{10} \alpha \left(E^\ast \xi \right)  $ \\
    $\alpha,\xi$&numerical constants, from ~\cite{kim2016}&$\alpha\simeq 1$,$\xi\simeq0.15$\\ $C$ &  brush elasticity (1/nm) & adjustable parameter \\
\end{tabular}\label{table:allpara}
\end{table}
\newline Once the effective volume fraction surpasses a higher critical value of $0.646$, the available free volume disappears, resulting in direct particle contacts and deformation~\cite{bernal1960packing,torquato2000random,ikeda2012unified,ikeda2013disentangling}. The microgel corona, which have a lower crosslinker content than the microgel cores are compressed. This corona compression and its impact on elasticity have been extensively studied~\cite{scheffold2010, romeo2013elasticity,bergman2018new, conley2019, scotti2020}. One of the approaches to model corona repulsive interactions is based on the scaling model for a polymer brush, originally proposed by Alexander and de Gennes and later refined~\cite{alexander1977adsorption,de1980conformations,zhulina1990theory}. It describes the repulsive force between two brush-coated surfaces, and it has been previously used to model the elasticity of the microgel corona in a good solvent \cite{scheffold2010, romeo2013elasticity, tu2007brush,conley2019, scotti2020}.  Here, we consider the mean field polymer theory treatment of the brush swelling and elasticity as summarized in reference \cite{espinosa2013impact}. In this approximation, the polymer brush assumes a parabolic density profile. Using the Derjaguin approximation, the force between two particles coated with a polymer brush can be expressed as follows:
\begin{equation}\label{eq:EspinosaForceChi}
 \begin{split}
        f(u)=-C(T)k_\text{B}T\frac{R_\text{core}}{L}\left[-\frac{1}{2u}-\frac{u^2}{2}+\frac{u^5}{10}+\frac{9}{10}\right] \end{split}\end{equation}
where $N_{p}$ is the degree of polymerization of the brush polymers,  with $C(T)=4\pi N_{p}^2 a^3/s^4 \times \tau\left(T\right)$.
\begin{figure*}[t!] 
\centering
\includegraphics[width=1\linewidth]{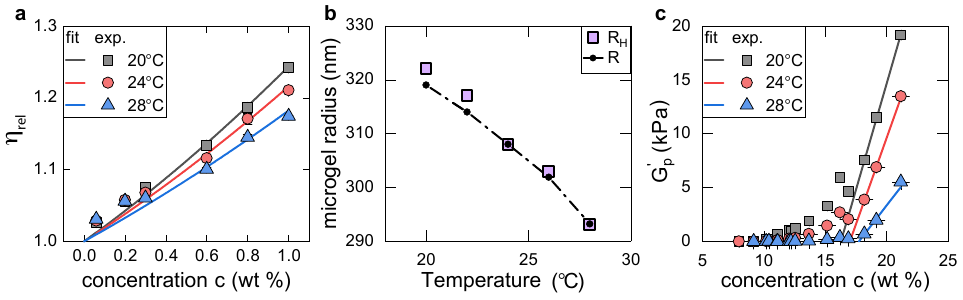}
\caption{Experimental characterization of the system parameters.
{\bf{a}} Relative viscosity of low-concentration microgel samples (represented by symbols). Data shown was measured at temperatures 20, 24, and 28 $^\circ$C with a rolling ball viscometer. The solid lines represent fits to the Einstein-Batchelor equation. {\bf{b}} Full symbols and dashed line: Microgel (effective) particle radius ($R$) as a function of temperature obtained from the swelling ratio $k$ and the number density $N/V$. Open squares indicate the hydrodynamic radius $R_\textrm{H}$ measured through Dynamic Light Scattering (DLS). {\bf{c}} Elastic modulus plateau values ($G^{'}_{p}$) obtained through rheometry (represented by symbols), along with corresponding fits to the linear regime at the highest concentrations (solid lines).}\label{fig:exp}
\end{figure*}
By integration, we convert the force to energy per particle and obtain:
\begin{equation}\label{eq:EspinosaEnergyChi}
 \begin{split}\frac{F_\text{brush}}{N k_\text{B}T}=
C(T)\frac{R_\text{core}}{30 }  \left(u^6-10 u^3+54 u-30 \ln (u)-45\right)
\end{split}\end{equation} for $u=h/2L\le1$.
The parameter $h$ represents the distance between the surfaces of the microgel cores, while $L$ stands for the equilibrium thickness of the polymer brush. The average distance between the polymer chains anchored on the microgel core is represented by $s$, and $a$ is the typical size of a polymer molecule. Although water is a good solvent for pNIPAM microgels well below $T_\textrm{LCST}$, its quality deteriorates as $T$ approaches $T_\textrm{LCST}$ \cite{espinosa2013impact,camerin2018modelling,tavagnacco2021molecular}. This effect is captured by the dimensionless virial coefficient $\tau \left(T\right)= T_\text{LCST}/T - 1 \simeq 1- T/T_\text{LCST} $ where $\tau \left(T\right) \propto A_2\left(T\right)$  is proportional to the experimentally accessible second virial coefficient. The linear decay of  $A_2\left(T\right)$ was revealed experimentally for the single chain phase transition of pNIPAM by Kubota et al. ~\cite{kubota1990single}, based on light scattering.

The relationship between $h$, the core volume fraction $\zeta$, and the deformation volume fraction $\zeta_d$ was established by Kim et al. \cite{kim2016}.
\begin{equation}\label{eq:3}
    \frac{h}{R_\textrm{core}}\simeq2\cdot\left(0.646\right)^{1/3}  \left[\zeta^{-1/3}-(0.646+\zeta_\textrm{d} -\alpha\gamma^2)^{-1/3}\right]\\
\end{equation}
where $0.646$ is the jamming volume fraction of disordered monodisperse spheres.
In summary, we model corona interactions by employing a polymer brush model that considers the influence of solvent quality. Anticipated is a diminishing strength of brush repulsion as we approach the lower critical solution temperature ($T_\text{LCST}$).
\newline When the particle concentration is high, the microgel corona becomes fully compressed onto the densely crosslinked core. Upon further increase of the microgel particle number density $\frac{N}{V}$, the microgel cores must undergo additional adaptation, as demonstrated in various studies~\cite{scheffold2010, romeo2013elasticity, bergman2018new, gnan2017silico, rovigatti2017internal, rovigatti2019connecting, tavagnacco2021molecular}. This adaptation results in deformation, faceting, and compression of the core, creating space for additional microgels. As observed in superresolution microscopy and neutron scattering studies, the microgel size decreases, indicating a volume change due to the softness of its core~\cite{gasser2014form,conley2017}. Seth, Cloitre, and Bonnecaze showed that the contact elastic modulus $E^\ast$ of touching spheres can be related to the Laplace pressure in jammed emulsion droplets as $E^\ast\simeq 10 \sigma/R$, where $\sigma$ represents the surface tension of the droplets~\cite{seth2006elastic}. The contact modulus $E^\ast$ is further related to the Young’s (bulk) modulus $E$ and the Poisson’s ratio $\nu$ through $E^\ast=E/\left[2\left(1-\nu^2\right)\right]$~~\cite{seth2006elastic,houston2022resolving}.  Using the results of ref.~\cite{mason2014crossover}, we express the leading-order term in the free energy associated with particle core deformation as follows:
\begin{equation}\label{eq:4}
\frac{F_\textrm{core}}{N}=4\pi \frac{E^\ast}{10} \xi R_\textrm{core}^{3}\zeta^{2}_\textrm{d}
\end{equation} 
where $\xi$ is a dimensionless geometric parameter considering the entire disordered system. It encompasses the variously sized facets and different coordination numbers within the distribution of all particles.
\newline The total free energy can now be written as:
\begin{equation}\label{eq:5}
F_\textrm{tot}=F_\textrm{ent}+F_\textrm{brush}+F_\textrm{core}
\end{equation}
where $F_\textrm{ent}$ denotes the entropic translational free energy $
F_\textrm{ent}=-3 N k_\textrm{B}T\ln(0.646+\zeta_d-\zeta-\alpha\gamma^2)$~\cite{kim2016}. 
We derive the plateau shear modulus by taking the derivative of the total free energy, Eq.~\eqref{eq:7}, under the minimization condition $\zeta_d=\zeta^{*}_d$ as explained in \cite{mason2014crossover} and analyze the shear modulus for different concentrations and various temperatures.
\newline It is helpful and informative to assess the plateau shear modulus under conditions where the corona term is omitted and $k_\text{B}T$ approaches zero. According to findings in \cite{mason2014crossover}, substituting $\sigma/R$ by $E^\ast/10$ the corresponding formula for the shear modulus is:
\begin{equation}\label{eq:modhighc}
G^\prime_p\left(\zeta\right) = \frac{12}{10} \alpha E^\ast \xi \times \zeta \left(\zeta-0.646\right)
\end{equation}
This expression is expected to be the prevailing term for $\zeta>0.646$. In the following, based on previous emulsion studies~\cite{mason2014crossover,kim2016}, we set $\alpha=1$ and $\xi= 0.15$ and consider $E^\ast$ an effective core contact elasticity parameter.
\newline We note that in reference~\cite{ghosh2019linear} Ewoldt, Schweizer, and coworkers reported on experiments and a model for the rheology of soft, repulsive pNIPAM microgel suspensions. Their study is distinct from our approach in several ways. They investigate self-crosslinked pNIPAM microgels, which may lead to a different core-shell architecture compared to the chemically crosslinked microgels considered here. Additionally, they study suspensions under high swelling at 10$^\circ$C, far from the volume phase transition. In contrast to our approach, they largely exclude the 'soft jammed' scenario, where microgels are considered soft particle entities that come into literal contact. In this jamming scenario, particles may deform, form facets, and store elastic energy in a granular manner. Instead, they explore a Brownian glassy suspension scenario for the entire concentration regime without invoking athermal soft jamming. However, as shown earlier~\cite{conley2017,conley2019}, the particle nature of BIS-crosslinked microgels is evident from superresolution microscopy, and the rheological properties are consistent with this picture concerning the onset of elasticity. Faceting and interfacial contacts can explain the absence or presence of microgel interfacial friction at low or high microgel concentrations~\cite{conley2019}.
\newline Their dynamical model invokes a tagged particle in a liquid using a generalized Langevin equation within the framework of the naive ideal Mode-Coupling Theory (NMCT) of single-particle dynamics developed by Schweizer~\cite{schweizer2005derivation}. While their method for finding the rheological properties involves a qualitatively different 'dynamic' approach compared to our quasi-free energy model, the fact that they define explicit particle-particle interaction potentials is similar. It would be interesting in future work to compare the predictions of both models using the same microscopic details to determine which model makes better predictions.

\section*{RESULTS AND DISCUSSION}
Our model possesses a significant advantage in that we can independently determine all parameters through experiments, except for the corona elasticity parameter $C(T)$. In particular, we derive the radius of swollen microgels, and consequently the effective volume fraction, using a combination of rheometry and microscopy—an approach that, to our knowledge, has not been attempted before. This all-rheology approach has the advantage of not relying on additional model assumptions typically made in scattering experiments, such as the equivalence of the radius in the Stokes-Einstein and Einstein-Batchelor relations, or the core-shell architecture in the popular fuzzy sphere model~\cite{stieger2004small,ScheffoldSM24}. To enhance clarity, we have included Table~\ref{table:allpara}, which provides details on all model parameters along with information on how each is determined. In the following sections, we will elucidate our step-by-step approach. Additionally, it is worth noting that independent studies indicate that $C(T)$ is expected to linearly decay with temperature \cite{kubota1990single}, reaching zero as it approaches $T_\text{LCST}$. Furthermore, it is anticipated that $E^\ast\left(T\right)$ will remain approximately constant, as suggested by previous research \cite{voudouris2013micromechanics}. We will leverage this information to validate our model assumptions.
\begin{table*}[t!]\centering
\caption{Table of parameters utilized to characterize the experimental data presented in Fig.~\ref{fig:gpfit}, acquired as outlined in the text.}
\resizebox{\textwidth}{!}{
\begin{tabular}{r|rrrrrrrrr}
Temperature ($^\circ$C) 	& 20 & 21 & 22 & 23 & 24 & 25 & 26 & 27 & 28\\
\hline
$R$ (nm) & 319.0&316.6	&314.0&	311.0&	308.0	&305.0&	301.9	&297.6	&293.2\\
$R_\textrm{core}$  (nm)  &  254.5 & 253.1 & 251.5 &  250.0 &  248.7 & 247.3 & 246.1 & 244.9 & 243.6\\
$L=R-R_\textrm{core}$ (nm)& 64.5 & 63.5 & 62.5 & 61.0 & 59.3 & 57.7 & 55.9 & 52.7 & 49.6 \\
\hline 
$k$  & 8.178 & 7.989 & 7.797 & 7.579 & 7.360 & 7.145 & 6.930 & 6.640 & 6.350\\
$k_\textrm{core}$ & 4.150 & 4.081 & 4.004 & 3.935 & 3.874 & 3.811 & 3.752 & 3.698 & 3.641\\
\hline
$E^{\ast}$ (N/m$^2$) & 526183 & 534233 & 544965 & 544188 & 528307 & 502319 & 460258 & 397696 & 318656\\
$C \left(\text{1/nm}\right)$  & 30.08 & 26.48 & 23.17 & 20.06 & 16.24 &  11.37 & 7.48 & 4.58 & 1.83 \\
\label{sitab:fitparamLFfirst}
\end{tabular}}
\end{table*}
Initially, we determine the microgel concentration, denoted as $c$, in weight percent (wt $\%$) by drying and weighing a specific amount of a diluted stock suspension. Subsequently, viscosity measurements are utilized to establish a relationship between $c$ and the effective volume fraction $\phi$ of microgel samples, as referenced in previous works~\cite{senff1999temperature,scotti2020,conley2017,mohanty2017interpenetration,bergman2018new}. When the effective volume fractions are low, the relative viscosity (denoted as $\eta_\textrm{rel}=\eta/\eta_h$) of microgel samples adheres to the Einstein-Batchelor equation with $\phi=k c$ \cite{batchelor1977effect,brady1995normal}. In this scenario, as per Brady and Vicic's formulation, the expression for relative viscosity increase is given by $\eta_\textrm{rel}=\eta/\eta_h=1+2.5kc+5.9(kc)^2$, where $\eta_h$ represents the viscosity of the surrounding medium, specifically water. By measuring a series of samples with a known concentration ($c$ in wt $\%$) and analyzing $\eta_\textrm{rel}$, we can determine the temperature-dependent shift factor, referred to as the swelling ratio $k$ (see Fig.~\ref{fig:exp}a). We determined the number density of particles $N/V$ at a given $c$ by imaging a known volume $V$ using confocal microscopy and counting the particles therein. With knowledge of the number density, we can extract the effective radius $R$ shown in Fig.~\ref{fig:exp}b. The values obtained through this analysis compare well to the independently determined hydrodynamic radii obtained through dynamic light scattering, Fig.~\ref{fig:exp}b. 
\begin{figure*}[t!] 
\centering
\includegraphics[width=1\linewidth]{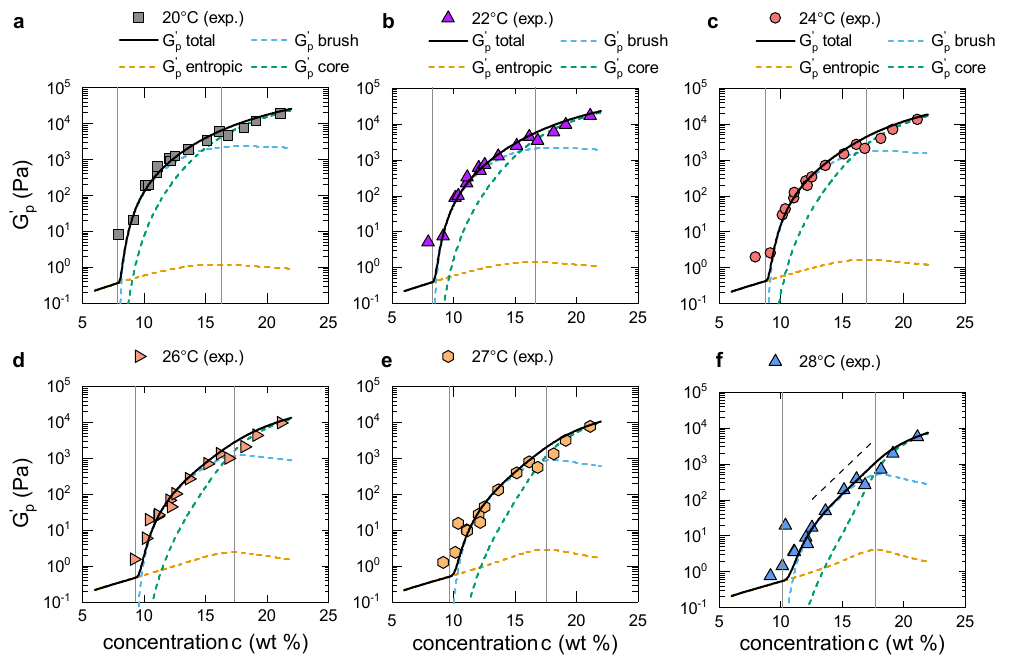}
\caption{Global model fit to the experimental  $G^\prime_p\left(c,T\right)$-data of the concentration-dependent shear modulus at various temperatures from  {\bf{a}} $\left(20^\circ\right)$ to  {\bf{f}} $\left(28^\circ\right)$. The elastic response's relative contribution from each of the three model terms is depicted individually as dashed colored lines. The effective volume fractions $\phi$ based on the unperturbed microgel radius measured in the dilute limit can be calculated by $\phi=k c$ using the $k-$values reported in Table \ref{sitab:fitparamLFfirst}. The black dashed line in panel {\bf{f}} indicates an exponential rise of the modulus at intermediate concentrations.}
\label{fig:gpfit}
\end{figure*}
We use oscillatory shear measurements in a cone-plate configuration to determine the linear elastic shear modulus in the high-concentration regime, Figs.~\ref{fig:exp} and~\ref{fig:gpfit}. In this context, we expect $G^\prime_p\left(c\right) = \frac{12}{10} \alpha \left(E^\ast \xi \right) \times \zeta\left(c\right) \left(\zeta\left(c\right)-0.646\right)$ with $\zeta(c)=c \cdot k_\text{core}$. By fitting the highest concentration data points, we obtain $k_\text{core}$ and $E^\ast$. We have verified that the fit is stable to within a few percent by varying the number of points included in the analysis. With the knowledge of $k_\text{core}$ and the number density $N/V$, we calculate the radius of the undeformed core $R_\text{core}$. We thus obtain the equilibrium thickness of the microgel corona via $L=R-R_\text{core}$.  

\begin{figure}[t!] \centering
\includegraphics[width=.53\linewidth]{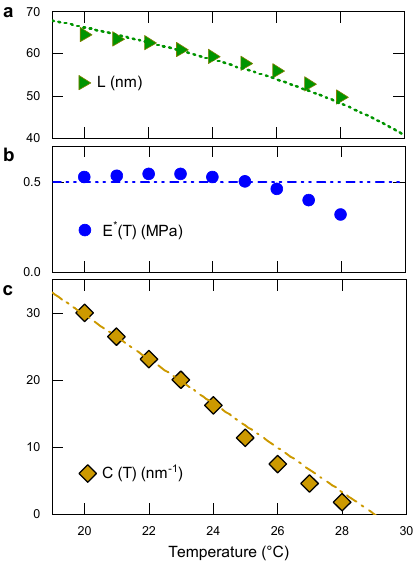}
\caption{Temperature dependence of the model's parameters. {\bf{a}} Equilibrium brush thickness $L=R-R_\text{core}$ (symbols). The dotted  line shows the prediction by the mean-field model: $L=190$~nm $\times \left(1- T/\left(306.15K\right)\right)^{1/3} $ where the temperature $T$ is expressed in Kelvin and the prefactor was adjusted for a best fit. {\bf{b}} Contact elastic modulus $E^\ast$ from a linear fit to $G^\prime_p\left(c\right)$ in the high concentration regime. The  line indicates a typical value of $E^\ast=0.5$MPa. {\bf{c}} Brush elasticity parameter in the mean-field-description, \eqref{eq:EspinosaEnergyChi}, obtained from a best global fit to the experimental data for $G^\prime_p\left(c\right)$. The dash dotted line shows a linear fit to the data with $C(T)=1000/$nm$\times \left(1- T/\left(302.15K\right)\right)$.}
\label{fig:fitparam}
\end{figure}

As an initial verification of our model assumptions, we analyze the experimentally determined temperature dependence of the brush thickness. The equilibrium thickness of the brush in the mean-field approximation is given by $L\sim N_p a \left(a/s\right)^{2/3} \tau\left(T\right)^{1/3}$, where, once again, $\tau \left(T\right)\simeq 1- T/T_\text{LCST}$. In Fig.~\ref{fig:fitparam}a, we compare the experimental data with the theoretical prediction and observe excellent agreement for $T_\text{LCST}\sim 33^\circ$C and $N_p a \left(a/s\right)^{2/3}\sim 190$~nm. With $a \left(a/s\right)^{2/3}$ on the order of $0.1$~nm (for $a\sim 0.5$~nm and $s\sim 5$~nm), this suggests that the corona consists of chains with approximately $N_p \sim 1800$ monomer units. Here, we have chosen $s$ to be comparable to the mesh size or correlation length of the cross-linked microgel core, on the order of a few nanometers~\cite{lietor2010structural}. 
\newline Returning to the analysis of the storage modulus, we note that the only remaining unknown parameter in the model for $G^\prime_p\left(c,T\right)$ is the parameter $C(T)=4\pi N_{p}^2 a^3/s^4 \times \tau\left(T\right)$, which determines the elasticity of the brush-like corona. For each temperature, we fit the model to the experimental data for $G^\prime_p\left(c,T\right)$ across the entire concentration range by adjusting $C(T)$ and find excellent agreement as shown in Fig.~\ref{fig:gpfit}. The values obtained for $C(T)$ at all temperatures are listed in Table~\ref{sitab:fitparamLFfirst}. Remarkably, we can reproduce the transitional single-exponential scaling of the shear modulus at the highest temperatures. This behavior, now understood to arise from the coupling between the elastic contribution of the coronas and that of the core, is represented by the black dashed line in Fig.~\ref{fig:gpfit}f. The relative contribution of each of the three model terms to the elastic response is depicted individually as dashed colored lines.

In Fig.~\ref{fig:fitparam}b and c, we present the temperature dependence of $E^\ast\left(T\right)$ and $C(T)$, respectively. As anticipated by polymer theory, the brush interactions soften as the solvent quality decreases when approaching the lower critical solution temperature (LCST) $T_\text{LCST}$\cite{espinosa2013impact}. The parameter $C(T)$ decays linearly following $\tau \left(T\right)\simeq 1- T/T_\text{c}$, where $T_\text{c}\sim 29^\circ$C is found to be slightly smaller than $T_\text{LCST}\sim 33^\circ$C. The prefactor $C(T)/\tau\left(T\right)=1000/$nm is consistent with the same set of brush parameters discussed earlier: $N_p\sim 1800$, $a=0.5$nm, and $s=5$nm. With these numbers, $4\pi N_{p}^2 a^3/s^4 \simeq 500/$nm. Despite minor disagreements regarding the transition temperature in the fit, our model provides an excellent description of the shear elasticity of microgel suspensions and pastes from the low to high concentration regime, with good qualitative agreement with the molecular properties of the polymer microgels. A key finding of our work is that we can relate microscopic microgel properties, such as the temperature-dependent polymer virial coefficient and corresponding brush elasticity, to the parameters of the rheological model. Such detailed understanding of rheological properties in terms of the characteristics of the building blocks is unusual for a system this complex. While the results derived from rheology and modeling are consistent, independent validation would be beneficial to further corroborate our findings. Indeed, recent optical tweezer measurements on similar but slightly larger microgels report intrinsic interaction potentials between colloidal particles in the weakly repulsive regime  for energies up to 10~k$_B$T~\cite{Muneton2024,zhang2024determining}. 
The overall agreement between direct measurements and the rheological analysis in the present work is fairly good, and the overall trend in terms of the decay of the virial coefficient is essentially the same in both experiments. Clearly, the quasi-equilibrium between core and corona deformation is established differently in a jammed packing compared to pairwise compression which can affect the absolute value of $C(T)$. 

\section*{CONCLUSION}
In summary, we present a free energy model describing the linear elastic shear modulus 
$G^\prime_p\left(c,T\right)$ of concentrated microgel suspensions and pastes. In our approach, we minimize the free energy by considering particle interactions, accounting for the vanishing free volume entropic contribution, the compression of a soft polymer corona, and a homogeneous polymer gel core. The explicit treatment of the corona in the polymer brush framework is a cornerstone of our model. It enables us to quantify the onset of elasticity, and the brush parameters we find can be directly related to the microgel corona swelling curve $L\left(T\right)$ , showcasing the consistency of the model. Notably, we find that the temperature dependence is well captured by the dimensionless virial coefficient, which quantifies the solvent quality for the microgel. In the present work, we focus on repulsive interactions below the volume phase transition temperature. However, it would be straightforward to incorporate weak attractive interactions into the model, and we may explore this in future work. A distinctive aspect of our approach is that we determine all parameters describing the system from oscillatory shear measurements, eliminating the need for indirect methods such as density profiles obtained from scattering or microscopy, which may not necessarily be representative for modeling the mechanical response. Our hierarchical model predicts constitutive properties of suspensions across the entire range of concentrations and temperatures studied. It unravels the contributions from various mechanical modes, including the reduced free volume, the microgel corona, and the core, providing insights into particle behavior under external stress. This model possesses predictive power for calculating the elastic response for soft repulsive interactions when the microgels are in a swollen state. Thus, it contributes to advancing our comprehension and utilization of polymer microgels in various fields.


\providecommand{\latin}[1]{#1}
\makeatletter
\providecommand{\doi}
  {\begingroup\let\do\@makeother\dospecials
  \catcode`\{=1 \catcode`\}=2 \doi@aux}
\providecommand{\doi@aux}[1]{\endgroup\texttt{#1}}
\makeatother
\providecommand*\mcitethebibliography{\thebibliography}
\csname @ifundefined\endcsname{endmcitethebibliography}
  {\let\endmcitethebibliography\endthebibliography}{}




\section*{Data availability}
All experimental data and model output discussed in the manuscript have been deposited in the the repository Zenodo (\href{https://zenodo.org/records/14771975}{14771975}).  All additional data sets generated during and/or analyzed during the current study are available from the corresponding author upon reasonable request.

\section*{Code availability}
All necessary details to perform the model calculations are provided both here and in our previous work~\cite{mason2014crossover}. However, the Mathematica (Wolfram Research, Inc., USA) implementation of the model developed for this study is proprietary and, therefore, not publicly available.

\section*{Competing interest}
The authors declare no competing interest.

\section*{Author Contributions} M.B. and Y.X.: contributed equally to this work. MJB, TGM and FS designed the study.  MJB performed most of the experiments and performed the initial data analysis. CZ contributed to the data analysis and carried out some additional experiments. FS and TGM derived the theoretical model. YX coded the model under the supervision of TGM. FS and MJB drafted the manuscript with contributions from all authors. All authors contributed to analyzing, interpreting, and revising the final version of the manuscript.

\section*{Acknowledgments}  
The Swiss National Science Foundation funded this project through project No. 188494. The project benefitted from support through  the National Center of Competence in Research Bio-Inspired Materials, Swiss National Science Foundation  project No. 205603. This work was supported by the University of California, Los Angeles. M.J.B. is grateful to Francois Lavergne for his assistance with rheology measurements.
 
\end{document}